# Ultralow-Crosstalk Silicon Electro-Optic Switch with Cascaded Phase Shifters for Loss Equivalence

Yating Wu and Tao Chu*

College of Information Science and Electronic Engineering, Zhejiang University, Hangzhou 310027, China
*Corresponding Author: chutao@zju.edu.cn

**ABSTRACT:** In silicon electro-optic (EO) Mach–Zehnder interferometer (MZI) switches, crosstalk is typically limited by beam imbalance between the MZI arms, primarily caused by the free carrier absorption loss during routing, thus hindering switch scalability. To address this issue, we propose a low-crosstalk push–pull EO MZI switch by cascading a lightly doped, long phase shifter ($PS_{LL}$) and a heavily doped, short phase shifter ($PS_{HS}$) to construct phase-shift arms. In both BAR and CROSS states, $PS_{LL}$ in one arm and $PS_{HS}$ in the other arm are simultaneously forward-biased, with $PS_{LL}$ provides a $\pi/2$ greater phase shift for switching, while $PS_{HS}$ balances loss of $PS_{LL}$, effectively minimizing crosstalk. Simulations indicate that the proposed switch achieves a crosstalk below −51 dB at a 1310 nm wavelength. The fabricated 2 × 2 silicon EO MZI switch exhibited crosstalk between −33 and −44.2 dB at 1316 nm, and maintained crosstalk below −30 dB across an impressive 61 nm optical bandwidth, with response times under 119 ns. Featuring single-pair electrode control, consistent two-state performance, and a compact size, this approach could enable high-radix switch fabrics in data centers and artificial intelligence compute clusters.

## 1. INTRODUCTION

Artificial intelligence (AI) and machine learning (ML) have recently driven a surge in data traffic across applications such as data centers and AI clusters.[1-5] To address this rapid growth, silicon photonics (SiPh) offers a promising solution for its high bandwidth, low power consumption, low latency, high integration density, and cost-effectiveness.[6-16] Among SiPh components, optical switches play a critical role in direct optical signal routing.[17-24] Mach–Zehnder interferometer (MZI) switches, in particular, are widely studied owing to their broad bandwidth and robustness against temperature or fabrication variations.[17, 20, 21] These switches are composed of beam splitters and phase-shift arms, with electro-optic (EO) phase shifter (PS) based on PIN junction[17,25,26] enabling rapid phase modulation through the free carrier dispersion (FCD) effect[27] in silicon material. However, carrier injection associated with FCD effect inherently induces free carrier absorption (FCA) loss[27], which exacerbates the loss imbalance between MZI arms,[17,28] with uneven beam splitters[28] being another minor factor. This imbalance causes switch crosstalk above −30 dB, which degrades signal routing quality, reduces system reliability, and restricts switch scalability. Consequently, developing ultralow-crosstalk silicon EO MZI switches are urgently required.

Various studies have been conducted to minimize FCA-induced loss imbalance between MZI arms. One common approach employs push–pull driving to adjust required arm phase difference for the CROSS/BAR state from $0/\pi$ to $\pm\pi/2$,[17,25,28] effectively reducing FCA loss. However, this method does not fully eliminate the loss disparities between MZI arms, leaving crosstalk above −30 dB. Another promising design replaces the conventional straight-line PS in the MZI switch with a nested EO MZI PS, providing a constant loss for both BAR and CROSS (B&C) states, which is balanced by incorporating either an attenuator or another MZI PH in the other arm.[29,30] This approach ensures consistent performance across B&C states but complicates the overall structure and control, with the need for at least two electrode pairs. A third novel method utilizes a self-heating EO PS in one MZI arm to counter FCA loss in the other arm, achieving low crosstalk in single-ended mode.[31,32] Nonetheless, this design inherently restricts response times to the microsecond scale due to intrinsic thermal effects, causes noticeable imbalances of switch losses between B&C states, and increases control complexity. To further suppress residual crosstalk in EO MZI switches, some studies have combined the above methods with techniques for uniform beam splitters such as employing large-tolerance couplers[28] or tunable splitters[31,32]. In addition to solutions based on simple 2 × 2 MZI fabric, another low-crosstalk strategy integrates switch units into a high-radix matrix switch,[33,34] utilizing only two inputs and outputs while directing crosstalk to idle ports. However, this approach significantly increases size, complicates control, and deteriorates performance such as switch loss.

In this study, we present an ultralow-crosstalk push–pull 2 × 2 silicon EO MZI switch, whose phase-shift arms comprise a lightly doped, long $PS_{LL}$ and a heavily doped, short $PS_{HS}$ with their equivalent diodes oriented oppositely. In both B&C states, $PS_{LL}$ in one MZI arm and $PS_{HS}$ in the other arm are simultaneously forward-biased; the $PS_{LL}$ provides a $\pi/2$ greater phase shift for switching, while the $PS_{HS}$ balances the $PS_{LL}$-induced loss to suppress crosstalk. Notably, this design features a simple single-electrode-pair control, consistent performance across B&C states, and a compact size via the single-stage MZI. Simulations showed switch crosstalk below −51.2 dB at the wavelength of 1310 nm. The fabricated switch achieved crosstalk between −33 to −44.2 dB at 1316 nm, and maintained crosstalk below −30 dB over an outstanding 61 nm bandwidth. Moreover, measured response times in both B&C states were under 119 ns under square wave signals. This design holds significant potential for large-scale signal routing in applications such as data centers, high-performance computers, and future AI compute clusters.

## 2. PRINCIPLE AND STRUCTURE

Figure 1a shows the schematic of the low-crosstalk push–pull 2 × 2 silicon EO MZI switch. The design includes 3 dB multimode interference (MMI) couplers, a $\pi/2$ phase-biased element, and phase-shift arms. The MMI couplers, based on paired interference principle,[35] ensure uniform and broadband splitting. The $\pi/2$ phase-biased element, designed by narrowing the waveguide,[17] is connected to the right-lower port of the first MMI coupler, counteracting the inherent $-\pi/2$ phase difference at MMI outputs. This allows the switch to achieve the BAR (CROSS) state with a $\pi/2$ ($-\pi/2$) phase difference between MZI phase-shift arms, enabling symmetric push–pull operation of PSs in both states. According to derivations in Section 6.1, switch crosstalk can be expressed as a function of the loss imbalance between MZI arms, which is defined as the absolute value of arm loss difference. Further, their relationship curve is plotted in Figure 1b, showing that crosstalk decreases as MZI arm losses equalize. Traditional push–pull designs modulate only one arm per state, resulting in significant arm loss imbalance and thus severe crosstalk. In contrast, this method modulates both MZI arms simultaneously, effectively balancing arm losses and achieving ultralow crosstalk.

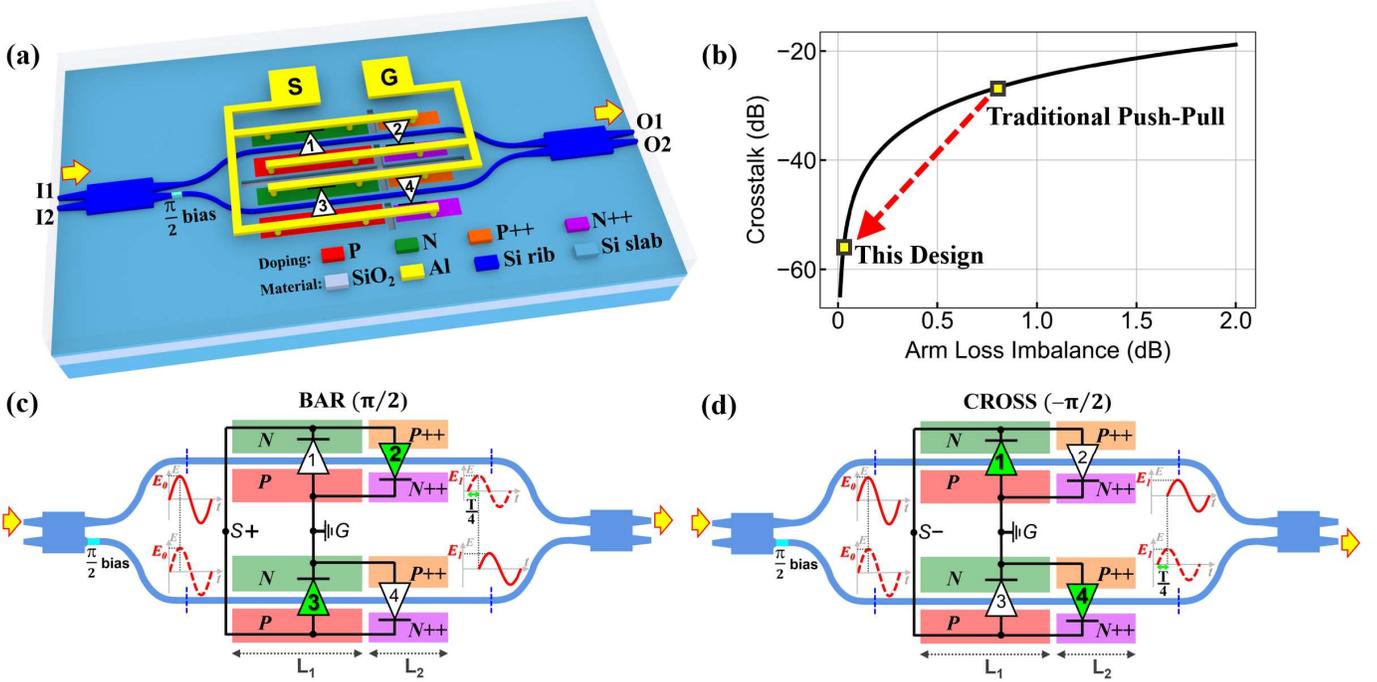

**Figure 1**. (a) Schematic of the low-crosstalk push-pull 2 × 2 electro-optic (EO) Mach–Zehnder interferometer (MZI) switch. (b) Crosstalk vs. loss imbalance between MZI arms for push–pull 2 × 2 MZI switches. Operation mechanism of the proposed switch in the (c) BAR state and (d) CROSS state.

Specifically, MZI arms in this design are configured with cascaded EO PSs in distinct configurations, as shown in Figure 1a. The upper (lower) phase-shift arm contains a lightly doped, long $PS_{LL1}$ ($PS_{LL3}$) and a heavily doped, short $PS_{HS2}$ ($PS_{HS4}$), each integrated with PIN junctions. Heavy doping for ohmic contacts is used in practice but omitted from the schematic for simplicity. Electrically, the PIN-based PS is typically modeled as a diode.[30] The configurations of EO PSs — including orientation of equivalent diode, doping concentration, doping spacing, PS length, and waveguide width — are optimized to enhance performance. As a result, $PS_{LL1}$ and $PS_{LL3}$ ($PS_{HS2}$ and $PS_{HS4}$) are identically configured with light (heavy) doping and a length $L_1$ ($L_2$), whereas $PS_{HS}$ and $PS_{LL}$ differ significantly, particularly in the opposite orientations of their equivalent diodes. Notably, all EO PSs share a single electrode pair, simplifying circuit control: inter-arm doping regions, including P-doped regions of $PS_{LL1}$ and $PS_{HS4}$, as well as the N-doped regions of $PS_{HS2}$ and $PS_{LL3}$, are connected to the ground (G) electrode, while the remaining regions are connected to the signal (S) electrode.

Figure 1c and 1d illustrate operation mechanisms of cascaded-PS phase-shift arms in the B&C states, respectively. For clarity, the electrical models of phase-shift arms are represented by black lines. In Figure 1c, when a positive voltage is applied to the signal electrode, both $PS_{HS2}$ and $PS_{LL3}$ are forward-biased, facilitating efficient phase modulation, with $PS_{LL3}$ contributing a larger phase shift. Simultaneously, $PS_{LL1}$ and $PS_{HS4}$ are reverse-biased with negligible modulation. At the voltage of $V_{\pi/2}$, $PS_{LL3}$ induces an additional $\pi/2$ phase shift than $PS_{HS2}$, achieving the BAR state. By optimizing the configurations of EO PSs, MZI arm losses, — primarily attributed to $PS_{HS2}$ and $PS_{LL3}$ — can be effectively equalized at $V_{\pi/2}$, thereby minimizing crosstalk. Conversely, in Figure 1d, applying a negative voltage to the signal electrode forward-biases $PS_{LL1}$ and $PS_{HS4}$, while $PS_{HS2}$ and $PS_{LL3}$ are reverse-biased. At $-V_{\pi/2}$, each PS operates in the opposite manner to its behavior in the BAR state, resulting in a $-\pi/2$ arm phase difference to achieve the CROSS state, as well as maintaining identical losses between MZI arms for low crosstalk.

## 3. SIMULATIONS

All devices operated in the O-band with transverse electric (TE) polarization. Simulations were performed on a silicon-on-insulator (SOI) platform with a 220-nm-thick top silicon layer, 3-μm-thick top and buried oxide layers, and a 130 nm etching depth. For PIN-based PSs, waveguide width and spacing between doping regions were set to 0.5 and 1.5 μm, respectively. According to specifications from Advanced Micro Foundry (AMF) company, $PS_{LL1\&3}$ were designed with lightly doped concentrations of 3.2 $e^{17}$ cm$^{-3}$ and 4.3 $e^{17}$ cm$^{-3}$ for P- and N-type regions. In contrast, $PS_{HS2\&4}$ featured heavily doped P++ and N++ regions with concentrations of 8.7 $e^{19}$ cm$^{-3}$ and 1.9 $e^{20}$ cm$^{-3}$. A constant doping approximation was utilized for simulations of EO PSs. More simulation details are provided in Section 6.3 and 6.4 of METHODS.

**3.1. Simulations for various $L_1/L_2$ combinations.** To determine the optimal lengths of $L_1$ and $L_2$ corresponding to $PS_{LL}$ and $PS_{HS}$ respectively, we analyze switch performance and obtain trends across 30 $L_1/L_2$ combinations at 1310 nm, with $L_1$ of 80, 100, 120, 140, and 160 μm, and $L_2$ ranging from 42 to 92 μm in 10 μm increments. Because of the symmetric push–pull behavior of PSs in B&C states, simulations were conducted only for the BAR state, as shown in Figure 2. Figure 2a summarizes the driving voltage $V_{\pi/2}$ required to achieve a π/2 phase difference between MZI arms for the BAR state. Figure 2b presents the loss imbalance between MZI arms at $V_{\pi/2}$. The minimum values were observed at $L_2$ of 42, 42, 52, and 52 (or 62) μm for $L_1$ of 100, 120, 140, and 160 μm, respectively, with corresponding loss imbalance of 0.101, 0.080, 0.007, and 0.129 (0.099) dB. Apparently, as $L_1/L_2$ deviated from these optimal combinations, arm loss imbalance increased, indicating a strong dependence on precise length combinations.

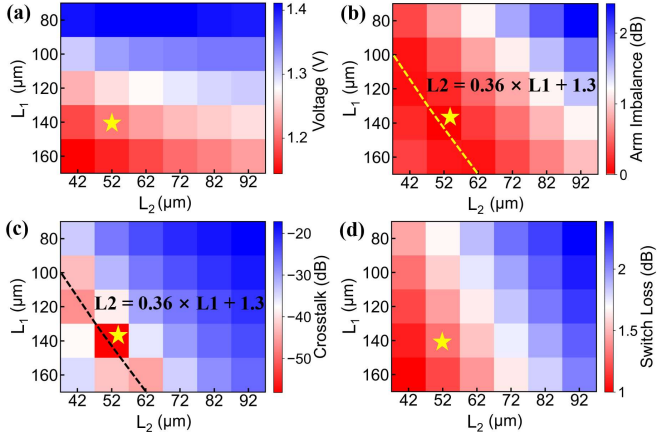

**Figure 2**. Simulation results for low-crosstalk EO MZI switches with various $L_1/L_2$ combinations: (a) Driving voltage $V_{\pi/2}$ to achieve the BAR state. (b) Arm loss imbalance at $V_{\pi/2}$, (c) switch crosstalk at $V_{\pi/2}$, and (d) switch loss at $V_{\pi/2}$.

To comprehensively analyze switch performance, additional simulations were conducted for passive components at 1310 nm. The simulations showed a 0.05 dB insertion loss, a 0.015 dB imbalance, and a −89.9° phase difference between outputs of MMI couplers[35]. The phase-biased element exhibited a bias of 89.8°. By combined these results with those of EO PSs, the overall switch performance was evaluated. Figure 2c illustrates switch crosstalk in the BAR state, defined as the ratio of leaked power from a non-ideal input port to the ideal output power along the routing path. For $L_1$ values of 100, 120, 140, and 160 μm, the lowest crosstalk occurred at specific $L_2$ values that minimize loss imbalance between MZI arms, as predicted in Figure 1b. Corresponding crosstalk values were −43, −46.6, −58, −42.4, and −44.9 dB, respectively. This trend revealed a positive correlation between optimal $L_1$ and $L_2$ values, approximately fitting the empirical formula $L_2 = 0.36 \times L_1 + 1.3$. Figure 2d summarizes switch losses at $V_{\pi/2}$ for all $L_1/L_2$ combinations, which remained below 1.5 dB for several low-crosstalk cases. Table 1 provides detailed simulation results for optimal $L_1/L_2$ configurations. Among these, we selected the lowest-crosstalk switch for further analysis, with $L_1$ of 140 μm for lightly doped $PS_{LL1\&3}$ and $L_2$ of 52 μm for heavily doped $PS_{HS2\&4}$, marked with a pentagram in Figure 2.

**Table 1. Switch performance for optimal $L_1/L_2$ combinations.**

| $L_1$ (μm) | 100 | 120 | 140 | 160 | 160 |
|---|---|---|---|---|---|
| $L_2$ (μm) | 42 | 42 | 52 | 52 | 62 |
| Voltage (V) | 1.306 | 1.235 | 1.207 | 1.165 | 1.185 |
| MZI Arm Phase Difference (rad) | π/2 | | | | |
| MZI Arm Loss Imbalance (dB) | 0.101 | 0.080 | 0.007 | 0.129 | 0.099 |
| Crosstalk (dB) | −43 | −46.6 | −58 | −42.4 | −44.9 |
| Switch Loss (dB) | 1.31 | 1.2 | 1.31 | 1.2 | 1.4 |

### 3.2. Simulations with $L_1$ of 140 μm and $L_2$ of 52 μm.

Figure 3 illustrates voltage-dependent characteristics for PIN-based $PS_{LL}$ and $PS_{HS}$ using the optimal length of 140 and 52 μm, respectively. In Figure 3a, the complex effective refractive index ($n_{eff}$) of the TE0 mode in EO PSs was simulated as a function of the applied voltage. For lightly doped $PS_{LL}$, the real part of the refractive index (Real($n_{eff}$), black dashed curve) decreased monotonically with increasing voltage. In contrast, for heavily doped $PS_{HS}$, Real($n_{eff}$) (black solid) initially decreased but then increased as thermal effects gradually counteract FCD effect at higher voltages. The imaginary part of the refractive index (Imag($n_{eff}$), red curves) increased more steeply for $PS_{HS}$, indicating stronger absorption. Using the index data from Figure 3a, Figure 3b and 3c present the voltage-dependent curves for phase shifts and loss in $PS_{1,2,3,4}$ (black curves with distinct line styles), as well as in the upper and lower phase-shift arms (red curves). As discussed in Figure 1c, positive voltages forward-biased $PS_{HS2}$ and $PS_{LL3}$, inducing noticeable phase shifts and FCA loss in Figure 3b,c, with $PS_{LL3}$ showing a larger phase shift; simultaneously, $PS_{LL1}$ and $PS_{HS4}$ were reverse-biased with negligible phase shifts and extra loss. The opposite occurred under negative voltages. Finally, the phase difference and loss difference between MZI arms, defined as values of the upper arm minus those of the lower arm, were calculated and depicted in Figure 3d, represented by the black and red curves, respectively. At voltages of ± 1.207 V, arm phase difference approached ±π/2, corresponding to BAR or CROSS states, while arm loss differences were approximately ±0.007 dB, thereby suppressing crosstalk.

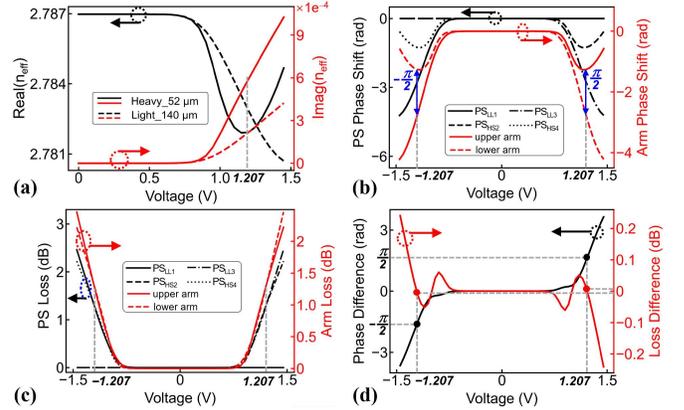

**Figure 3**. Voltage-dependent curves: (a) Complex effective refractive index of TE0 mode in EO $PS_{HS}$ and $PS_{LL}$. (b) Phase shift and (c) loss for $PS_{1,2,3,4}$ and MZI arms. (d) Phase difference and loss difference between MZI arms.

Based on the analysis of EO PSs, further simulations of the entire switch were performed, including voltage- and wavelength-dependent transmissions. Figure 4a and 4b present the voltage–optical transmission (V−T) curves for the proposed and conventional push–pull EO MZI switches with the same total modulation length of 192 μm, respectively. In the 2 × 2 switch, input and output ports were labeled as $I_{1,2}$ and $O_{1,2}$, with solid and dashed curves denoting the transmission for paths I1–O1 and I2–O1. As shown in Figure 4a, the proposed switch achieved the BAR (CROSS) state at a voltage $V_{BAR}$ ($V_{CROSS}$) of 1.207 V (−1.206 V), with crosstalk of −58 dB (−51.2 dB) and switch losses of 1.31 dB. The slight deviation of $V_{CROSS}$ from −1.207 V was attributed to minor imperfection in phase compensation by the phase-biased element for the MMI, resulting in a small disparity in two-state crosstalk. In comparison, in Figure 4b, conventional single-doping push–pull switches exhibited crosstalk below −28 dB for heavy doping (red) and below −29.2 dB for light doping (black) in both B&C states, with arm loss imbalances being 0.67 and 0.62 dB, respectively. These results demonstrated that the proposed design improved crosstalk by more than 20 dB.

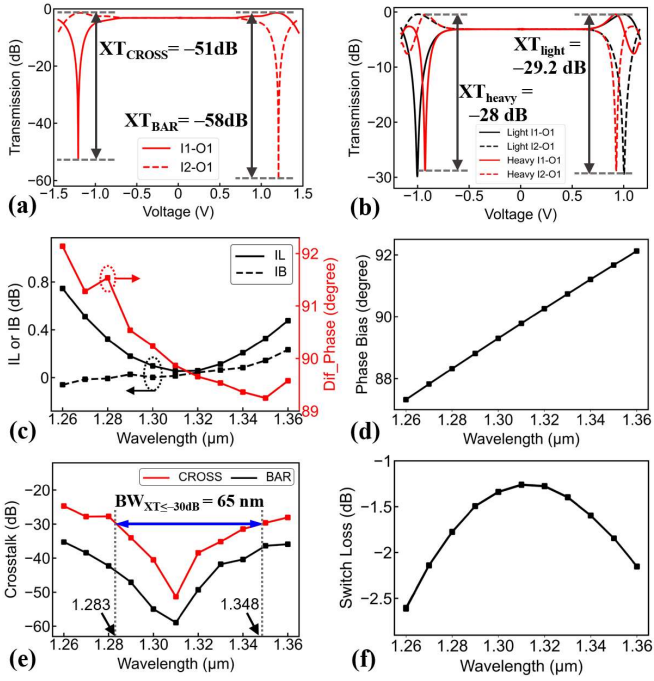

**Figure 4**. Transmission vs. voltage for the (a) proposed and (b) traditional switches, both with a 192 μm total modulation length in push–pull driving. (c) Insertion loss (IL), imbalance (IB), and phase difference (Dif_Phase) between outputs for multimode interference (MMI) couplers vs. wavelength. (d) Phase biases of π/2 phase-biased element vs. wavelength. (e) Switch crosstalk and (f) switch loss vs. wavelength at fixed $V_{BAR}$ or $V_{CROSS}$.

To evaluate the spectral performance of the proposed switch, broadband characteristics of the MMI coupler and π/2 phase-biased element were simulated. Figure 4c shows that across the 1270–1360 nm wavelength range, the MMI coupler exhibited insertion losses (IL; black solid) below 0.5 dB, imbalances (IB; black dashed) under 0.23 dB, and phase difference deviations at outputs (red solid) within 1.3° from the ideal 90°. Figure 4d indicates that actual biases of the phase-biased element varied between 87.3° and 92.2° over a 100-nm bandwidth. Additionally, phase-shift arms with intrinsic wavelength insensitivity, maintained loss imbalance between MZI arms near 0.007 dB in the B&C states, contributing to the low theoretical crosstalk. Building on these results, Figure 4e and 4f present the switch crosstalk and insertion losses as functions of wavelength, with $V_{CROSS}$ at −1.206 V and $V_{BAR}$ at 1.207 V, as determined from the V–T curve in Figure 4a. Over a 100-nm bandwidth, switch crosstalk in both B&C states was below −25 dB, primarily constrained by the imperfect phase match. Specifically, at shorter or longer wavelengths, the phase-bias element either under-compensated or over-compensated the phase difference between MMI outputs. Consequently, as wavelength deviated from 1310 nm, the voltage corresponding to the lowest crosstalk gradually shifted from the set voltage $V_{CROSS}$ and $V_{BAR}$, leading to an increased crosstalk at these fixed voltages. Notably, among a 65 nm bandwidth (1283–1348 nm), crosstalk was below −30 dB. In addition, switch losses retained under 2.7 dB across the 100 nm range, attributed to combined impacts of MMI loss and FCA losses. Supporting Information Section S1 and S2 further demonstrate the superior robustness of this design against manufacturing variations. For global waveguide width deviations within ± 40 nm, crosstalk remained below −46.6 dB. Similarly, for doping concentration variations in the $PS_{LL}$ or $PS_{HS}$ within ± 50 % of nominal designed values, crosstalk was below −39.7 dB.

## 4. FABRICATION AND MEASUREMENTS

**4.1. Fabrication.** The low-crosstalk 2 × 2 EO MZI switch was fabricated on a 220-nm SOI platform with 3-μm-thick claddings, using AMF standard process. Figure 5a shows the micrograph of the overall switch, while Figures 5b and 5c provide detailed views of the first MMI coupler integrated with a π/2 phase-biased element, and phase-shift arms with cascaded PIN PSs, respectively. The device featured a 130-nm etching depth and a 500-nm phase-shift waveguide width. All doping regions were positioned 0.5 μm from the waveguide edge. $PS_{LL1}$ and $PS_{LL3}$ with a length of 140 μm, were lightly doped, while $PS_{HS2}$ and $PS_{HS4}$ with a length of 52 μm, were heavily doped. Thermal isolation was implemented between EO PSs, and grating couplers (GCs) were used for TE0 light coupling. The switch, excluding GCs and electrode pads, measured only 560 × 80 μm². For comparison, push–pull switches with single lightly or heavily doped 192-μm-long phase-shift arms were also fabricated, as illustrated in Figure 5d and 5e.

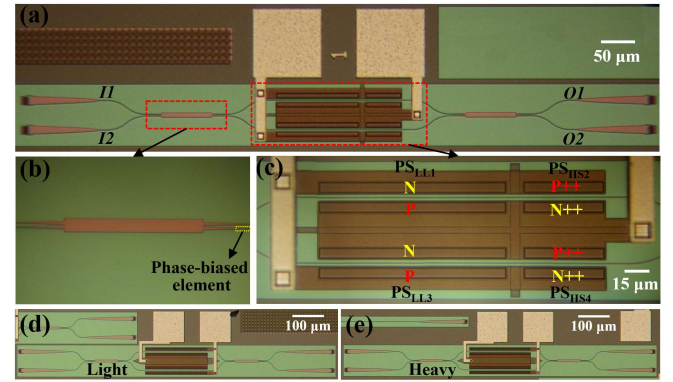

**Figure 5**. Micrographs of (a) the proposed low-crosstalk 2 × 2 EO MZI switch, (b) MMI coupler integrated with a π/2 phase-biased element, and (c) phase-shift arms with cascaded EO PSs. Micrographs of traditional push–pull switches with single (d) lightly or (e) heavily doped phase-shift arms.

**4.2. Static test.** Transmission characteristics of the proposed switch as a function of voltages were measured at 1316 nm using the experimental setup in Section 6.5. Figure 6a and 6b display V–T curves for traditional push–pull EO MZI switches with single lightly or heavily doped phase-shift arms, respectively. The lightly doped switch exhibited crosstalk of −24.3 and −26.5 dB in the BAR and CROSS states, while the heavily doped switch showed crosstalk of −23.4 and −18.9 dB in these states. Figure 6c presents V–T curves for the proposed switch, where black solid, red dashed, red solid, and black dashed lines represent paths I1–O1, I1–O2, I2–O2, and I2–O1, respectively. At a voltage of 1.49 V, the switch achieved the BAR state, routing signals from port I1 (I2) to O1 (O2), with switch crosstalk of −33 (−44.2) dB and switch losses below 2.6 dB. Likewise, the CROSS state was reached at −1.49 V, directing optical signals from ports I1 (I2) to O2 (O1), whose crosstalk was −38.5 (−33.7) dB and losses were below 2.5 dB. In summary, the proposed switch demonstrated notably lower crosstalk compared to both lightly or heavily doped traditional switches with the same modulation length, aligning closely with simulation results in Figure 4 a,b.

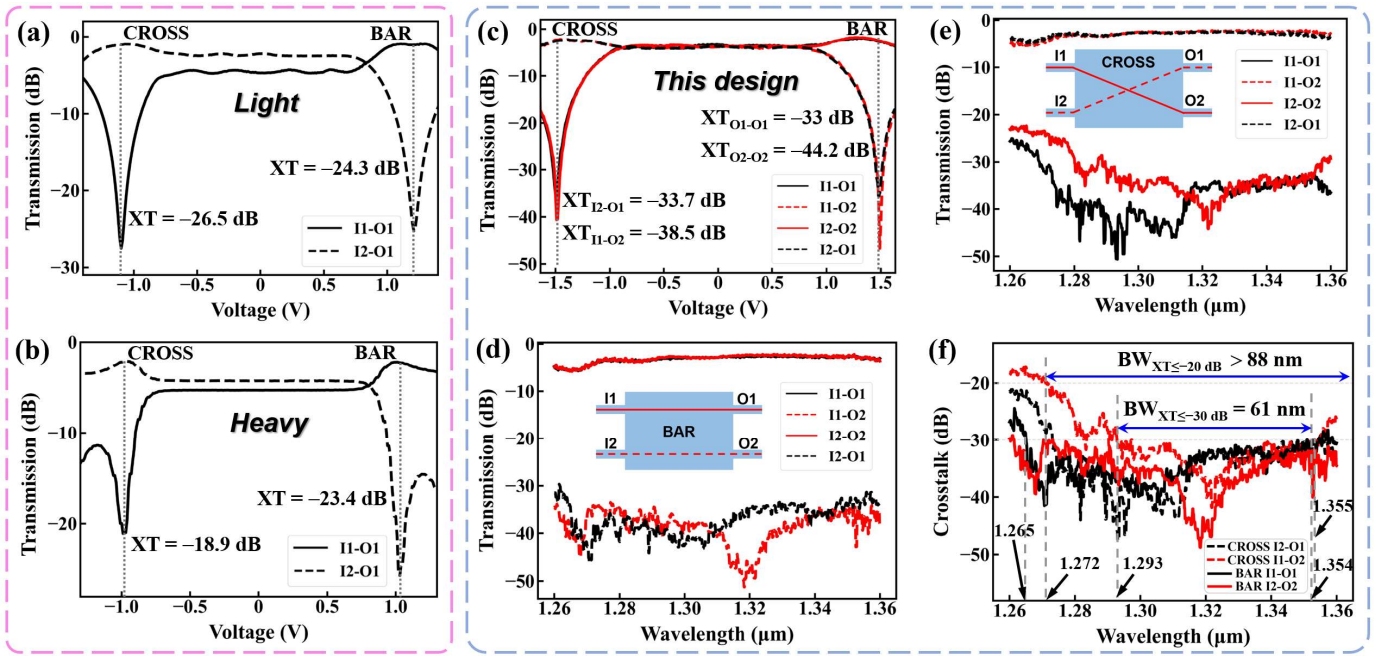

**Figure 6**. Measured transmission vs. voltage for traditional push–pull switches with (a) lightly or (b) heavily doped configurations. (c-f) Experimental results for the proposed low-crosstalk EO MZI switch: (c) Transmission vs. voltage. Transmission spectrum in the (d) BAR and (e) CROSS state. (f) Crosstalk spectrum.

For the proposed switch, transmission spectra in both B&C states were measured, as shown in Figure 6d and 6e. In the BAR state (Figure 6d) at 1.49 V, spectra for all paths were displayed. Based on these curves, crosstalk for routing paths I1–O1 and I2–O2 was calculated by subtracting the ideal output power (black/red solid) from the leaked power of the non-ideal input port (black/red dashed), with results plotted as solid black and red curves in Figure 6f. Notably, across a 100-nm bandwidth (1260–1360 nm), crosstalk for these routing paths remained below −24.5 dB, and within a narrower 89-nm range (1265–1354 nm), it stayed below −30 dB. In the CROSS state (Figure 6e) at −1.49 V, crosstalk for routing paths I2–O1 and I1–O2 was similarly calculated and shown as dashed curves in Figure 6f. For path I2–O1 (black dashed), crosstalk was below −21 dB across 1260–1360 nm, and for path I1–O2 (red dashed), it remained below −20 dB over an 88-nm range (1272–1360 nm). Additionally, within a 62-nm wavelength range from 1293 to 1355 nm, crosstalk for all paths in the CROSS states was below −30 dB. In summary, in both B&C states, crosstalk across all routing paths remained below −30 dB within a 61-nm wavelength range from 1293 to 1354 nm, while their insertion losses were measured as follows: 2.3–3.2 dB for path I1–O1, 2.1–3.0 dB for I2–O2 and I1–O2, and 2.3–3.4 dB for I2–O1. Moreover, crosstalk in both states stayed below −20 dB over an 88-nm bandwidth (1272–1360 nm), which was evaluated to extend into E-band by approximately 20 nm beyond 1360 nm according to trends in Figure 6f, resulting in an estimated total bandwidth of 108 nm for crosstalk below −20 dB. This broadband performance makes the switch suitable for integration with wavelength division multiplexing (WDM) technology, significantly enhancing routing capacity.

**4.3. Transient test.** The transient response of the proposed switch, defined as the time for the optical power transition from 10% to 90%, was measured at 1316 nm, as illustrated in Figure 7. The experimental setup is detailed in Section 6.5. A 50-kHz square wave signal (black lines) with switching times below 0.5 ns and amplitudes of ± 1.49 V was generated by an arbitrary waveform generator (AWG). During the CROSS-to-BAR transition (Figure 7a) where voltage varied from −1.49 to 1.49 V, response times for paths I1–O1 (red line) and I1–O2 (blue line) were 79 and 118 ns, respectively. Conversely, during the BAR-to-CROSS transition (Figure 7b), response times were 117 ns for paths I1–O1 and 87 ns for I1–O2. The switching speed of the proposed design was slightly slower compared to conventional EO switches[17, 29] due to minor thermal effect from $PS_{HS2\&4}$, as discussed in Section 3.2 for Figure 3a. Nevertheless, it remains well-suited for data routing and transfer in applications such as AI pods, data centers, optical sensing, and LiDAR, where switching times on the order of hundreds of nanoseconds are acceptable.[24] Additionally, the response curves in Figure 7 are non-monotonic, because simultaneously forward-biased $PS_{HS}$ and $PS_{LL}$ during switching, operate at different speeds, which results in a non-monotonic phase difference between MZI

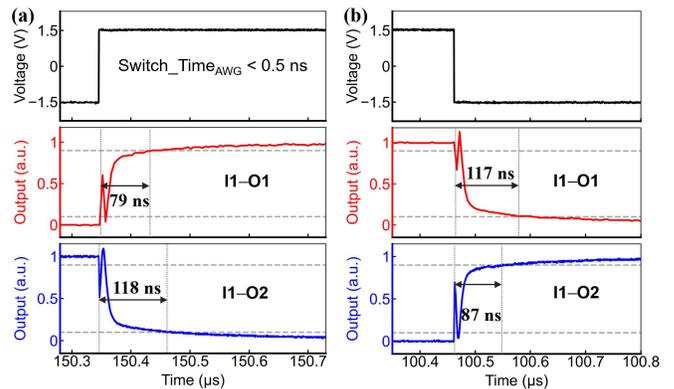

**Figure 7**. Response times of the proposed low-crosstalk switch: (a) CROSS-to-BAR and (b) BAR-to-CROSS transitions for paths I1–O1 and I1–O2.

**Table 2. Performance comparisons of low-crosstalk 2 × 2 silicon EO MZI switches.**

| Ref | Structure | Band | Modulation Length (μm) | Electrode Pairs | Best XT[D] (dB) | Worst XT (dB) within BW[D] (nm) | Switch Speed | Switch Loss (dB) |
|---|---|---|---|---|---|---|---|---|
| 33 | Four-element switch matrix | | 400 × 4 | 4 | [−35, −52] | −31 / 40 | NA[C] | ~8 |
| 30 | Balanced nested MZI[A] | | 250 | ≥2 | −50 (Sim.[E]) | NA[C] | NA[C] | NA[C] |
| 31 | Differential PS pair + two passive DC[A] | C | 500 | 2 | −60 (Sim.[E]) | −30 / 24 (Sim.[E]) | (a) 12 μs; (b) 20 ns by pre-emphasis and differential control (Sim.[E]) | < 2 (Sim.[E]) |
|  | Differential PS pair + two active DC[A] | | | 4 | −∞ (Sim.[E]) | −30 / 30 (Sim.[E]) | | < 2 (Sim.[E]) |
| 32 | Differential PS pair + variable splitter | | | ≥3 | −40 | NA[C] | (a) 10 μs; (b) 100 ns by differential control | NA[C] |
| 28 | Single-ended + DC coupler | O | 250 | ≥1 | [−21.1, ~−35] | −20 / 19 | ~4 ns | 1.6 |
|  | Single-ended + BBC[B] | | | | [−20.5, ~−25] | −20 / 30 | | 1.45 |
|  | Push pull + DC coupler | | | | [−27.1, ~−35] | −20 / 12 | | 1.2 |
|  | Push pull + BBC[B] | | | | [−23, ~−28] | −20 / 57; −23 / 46 | | 1 |
| 29 | Nested MZI switch | | | ≥3 | ~−34.5 | −20 / < 1.5; −30 / < 0.5 | < 4 ns | 2 |
| This work | Cascaded PSs + Push pull | | 192 | 1 | [−33, −44.2] | −30 / 61; −20 / >88; −20 / 108 (predicted). | < 119 ns | <2.6 |

A: Directional coupler (DC); B: Broadband coupler based two-section DC coupler (BBC); C: No Access (NA); D: Best crosstalk (XT) occurs at a single wavelength λ₁, where [A, B] indicates the XT range at λ₁; E: These are simulated results (sim.), where others are measured results.

arms and accordingly a non-monotonic variation in the optical power.

**4.4. Discussions.** The switch losses, arising from FCA effect in the phase-shift arms and 3 dB couplers, were measured higher than simulated values, where phase-shift arms are considered as the main reason. Due to the absence of an accurate ion implantation model from the foundry, we employ constant doping approximations for trend simulations of EO PSs. Deviations between designed and manufacturing doping level makes optimal $L_1/L_2$ configuration shift from the simulated lengths of 140/52 μm, thereby degrading performance of switch loss and crosstalk. Referring to trends under various length combinations in Figure 2c,d, extending $L_1$ or shortening $L_2$ could potentially reduce both switch losses and crosstalk for manufacturing in AMF. Moreover, this study applied specific light and heavy doping concentrations, consistent doping spacing between P- and N-type regions, uniform etching depth, and identical waveguide widths for $PS_{1,2,3,4}$. Future work will investigate these parameters as additional degrees of freedom, utilizing optimization algorithms such as ML to further minimize switch loss and crosstalk. In addition, losses from beam splitter could be reduced by optimizing the design of MMI couplers or exploring alternative adiabatic couplers.[36]

With respect to the response speed of the switch, shorter $PS_{HS}$ length would mitigate its thermal effect[31] and further reduce switching times, provided it is designed according to the empirical formula in Section 3.1. Moreover, electrical control techniques like pre-emphasis and differential control[37, 38] can be implemented to enhance response speed, though they increase control complexity. Furthermore, optimizing doping configurations by leveraging increased degrees of freedom could minimize thermal effects and enable response time on the nanosecond scale. Table 2 compares the performance of the proposed EO MZI switch with those reported in previous studies.

## 5. Conclusion

This study presents a low-crosstalk push–pull 2 × 2 electro-optic MZI switch, cascading a lightly doped, long $PS_{LL}$ and a heavily doped, short $PS_{HS}$ with oppositely oriented equivalent diodes to construct phase-shift arms. In both BAR and CROSS states, a $PS_{LL}$ in one MZI arm and a $PS_{HS}$ in the other arm are simultaneously forward-biased; the $PS_{LL}$ provides a larger π/2 phase shift for switching, while the $PS_{HS}$ counteracts the loss of $PS_{LL}$ to minimize crosstalk. Notably, this switch offers a simple control method with single electrode pairs, consistent two-state performance through the push–pull operation, and a compact single-stage MZI design. Simulated at 1310 nm showed switch crosstalk of −51.2 dB. Experimental results for the fabricated switch demonstrated crosstalk between −33 and −44.2 dB at 1316 nm and below −30 dB across a notable 61 nm bandwidth in both BAR and CROSS states. Additionally, crosstalk was predicted to remain below −20 dB over a 108 nm bandwidth, extending into the E-band and suitable for WDM technology. Response times were less than 88 ns for routing paths and 119 ns for crosstalk paths, under standard square wave signals. This design is promising for constructing large-scale low-crosstalk switch arrays, supporting high-quality signal routing in applications such as data centers and AI clusters.

## 6. Methods

**6.1. Calculation for a push–pull 2 × 2 MZI switch.** The transmission matrix[39] of the 2 × 2 push–pull MZI switch is derived by combining the matrices of MMI couplers, the π/2 phase-biased element, and phase-shift arms. The BAR state is formulated and presented as Equation (1), where $T_{MMI}$ denotes the transmission efficiency of MMI couplers with uniform splitting, and $T_1$ ($T_2$) represents the transmission coefficients of upper (lower) MZI arm, with an π/2 arm phase difference corresponding to the BAR state. Building on Equation (1), crosstalk (XT) is expressed as Equations (2), which is applicable to both B&C states. Loss imbalance between MZI arms is calculated as the absolute value of arm loss difference, as given by Equation (3). By combining Equations (2) and (3), the relationship between crosstalk and arm loss imbalance is derived and expressed as Equations (4).

$$M_{2\times2\_switch} = M_{MMI} \times M_{phase\_shift\_arms} \times M_{\pi/2\_phase\_bias} \times M_{MMI}$$

$$= \sqrt{\frac{T_{MMI}}{2}} \times \begin{bmatrix} 1 & j \\ j & 1 \end{bmatrix} \times \begin{bmatrix} \sqrt{T_1} & 0 \\ 0 & \sqrt{T_2}j \end{bmatrix} \times \begin{bmatrix} 1 & 0 \\ 0 & -j \end{bmatrix} \times \sqrt{\frac{P_{MMI}}{2}} \times \begin{bmatrix} 1 & j \\ j & 1 \end{bmatrix} \quad (1)$$

$$= \frac{P}{2} \begin{bmatrix} \sqrt{T_1} - \sqrt{T_2} & \sqrt{T_1}j + \sqrt{T_2}j \\ \sqrt{T_1}j + \sqrt{T_2}j & \sqrt{T_2} - \sqrt{T_1} \end{bmatrix}$$

$$XT = 20lg \left| 1 - \frac{2}{\sqrt{T_1/T_2} + 1} \right| \quad (2)$$

$$IB_{arm\_loss} = \left| 10lg\left(\frac{T_1}{T_2}\right) \right| \quad (3)$$

$$XT = 20lg \left( 1 - \frac{2}{10^{\frac{IB_{arm\_loss}}{20}} + 1} \right) \quad (4)$$

**6.2. Calculation for EO PSs and MZI phase-shift arms.** The phase shift and loss of the PIN PSs are calculated using Equations (5) and (6),[27] where λ(μm), $n_{eff}$, and L(μm) represent the operating wavelength, complex effective refractive index of the guided mode, and PS length, respectively. The total loss and phase shift in the upper MZI arm are determined by combining $PS_{LL1}$ and $PS_{HS2}$, while those in the lower arm are calculated by $PS_{LL3}$ and $PS_{HS4}$. Finally, phase difference and loss difference between MZI arms are obtained by subtracting the respective values of two arms.

$$Phase\_Shift = \frac{2\pi}{\lambda} \times Real(n_{eff}) \times L \quad (5)$$

$$Loss(dB) = 10 \times lg(e) \times \frac{4\pi Imag(n_{eff})}{\lambda} \times L \quad (6)$$

**6.3. Simulation for various $L_1/L_2$ combinations.** (A) Carrier distribution and thermal field in PIN PSs were first simulated across swept voltages using Lumerical Charge engine. The complex refractive index $n_{eff}$ of TE0 mode was then obtained using Lumerical FDE solver. (B) Subsequently, phase shifts of PSs and the phase difference between MZI arms were calculated using Equation (5), with the voltage $V_{\pi/2}$ required for a π/2 arm phase difference summarized in Figure 2a. (C) At $V_{\pi/2}$, the loss of PSs and loss imbalance between MZI arms were calculated using Equation (6), with results shown in Figure 2b. (D) Additionally, simulated results for MMI couplers and phase-biased element via Lumerical FDTD solver, along with phase-shift arms, were imported into the switch model in Lumerical Interconnect engine (matching the schematic in Figure 1a) to generate V–T curves of switches for each $L_1/L_2$ combination. Finally, switch crosstalk and losses were extracted and presented in Figure 2c and 2d.

**6.4. Simulation for $L_1/L_2$ of 140/52 μm.** We conducted FDTD simulations for the MMI coupler and the π/2 phase-biased element to analyze their broadband spectral characteristics and fabrication tolerance. Following steps (A)–(C) in Section 6.3, Figure 3 were obtained. Building on steps (A) and (D) in Section 6.3, we simulated the switch V–T curves across various wavelengths and process errors, further summarizing insertion losses and crosstalk, as shown in Figure 4 and in the Supporting Information.

**6.5. Experimental Setup.** In the static testing of optical switches, light from a tunable laser (Santec TSL-550; tuning range: 1260–1360 nm) was first polarized to TE polarization using an off-chip polarization controller (PC), then coupled into and out of the switch chip through GCs, and finally detected by an optical power meter (OPM; Yokogawa, AQ2211). A computer-controlled voltage source (Keysight B2901BL) was used to apply voltages to the switch via a direct current probe, while output power was recorded to derive V–T curves and identify voltages ($V_{BAR}$, $V_{CROSS}$) for B&C states. For broadband measurements, the voltage was fixed at either $V_{BAR}$ or $V_{CROSS}$, and laser wavelengths were swept while the output optical power was recorded. Measured results were normalized using reference GCs as a baseline.

In the transient testing system of switches, 50 kHz square wave signals with sub-500 ps switching times, generated by an AWG (Active technologies, AWG5062), were applied to the switch chip via a high-frequency probe. TE-polarized laser light was coupled into the chip, while the output power was amplified by a Praseodymium-Doped Fiber Amplifier (PDFA; FiberLabs, AMP-FL56XX-OB/OP) and directed to both OPM and an off-chip photodetector (PD; LSIHPD-A12) through a 1:99 beam splitter. The electrical signal from PD was monitored on an oscilloscope (ROHDE&SCHWARZ, FSVA40) to measure switching speed.

## 7. ACRONYM

electro-optic — EO; Mach–Zehnder interferometer — MZI; phase shifters — PS; artificial intelligence — AI; machine learning — ML; silicon photonics — SiPh; free carrier dispersion — FCD; free carrier absorption — FCA; BAR and CROSS — B&C; multimode interference — MMI; ground — G; signal — S; transverse electric — TE; silicon-on-insulator — SOI; Advanced Micro Foundry — AMF; the complex effective refractive index — $n_{eff}$; Voltage–optical transmission — V–T; insertion loss — IL; imbalance — IB; grating coupler — GC; wavelength division multiplexing — WDM; arbitrary waveform generator — AWG; crosstalk — XT; polarization controller — PC; optical power meter — OPM; Praseodymium-Doped Fiber Amplifier — PDFA; photodetector — PD.

## 8. ACKNOWLEDGEMENT

The authors thank Haozhe Sun and Hengwei Zhang for supports in fabrication and testing. This study was funded by the National Key Research and Development Program of China (2023YFB2806600).

## 9. SUPPORTING INFORMATION

Supporting Information Available: <Simulations for waveguide width errors; Simulations for doping deviations.> This material is available free of charge via the Internet at http://xxx.

**Notes**: The authors declare no conflicts of interest regarding this article.

## REFERENCES

(1) Fang, C.; Guo, S.; Wang, Z.; Huang, H.; Yao, H.; Liu, Y. Data-Driven Intelligent Future Network: Architecture, Use Cases, and Challenges. *IEEE Commun. Mag.* **2019**, *57* (7), 34-40.
(2) Yang, J.; Xiao, W.; Jiang, C.; Hossain, M. S.; Muhammad, G.; Amin, S. U. AI-Powered Green Cloud and Data Center. *IEEE Access* **2019**, *7*, 4195-4203.
(3) Khani, M.; Ghobadi, M.; Alizadeh, M.; Zhu, Z.; Glick, M.; Bergman, K.; Vahdat, A.; Klenk, B.; Ebrahimi, E. SiP-ML: High-Bandwidth Optical Network Interconnects for Machine Learning Training. In Proceedings of the 2021 ACM SIGCOMM 2021 Conference, 2021.
(4) Jouppi, N.; Kurian, G.; Li, S.; Ma, P.; Nagarajan, R.; Nai, L.; Patil, N.; Subramanian, S.; Swing, A.; Towles, B.; et al. TPU v4: An Optically

Reconfigurable Supercomputer for Machine Learning with Hardware Support for Embeddings. In Proceedings of the 50th Annual International Symposium on Computer Architecture, 2023.
(5) Baziana, P.; Drainakis, G.; Georgantas, D.; Bogris, A. AI and ML Applications Traffic: Designing Challenges for Performance Optimization of Optical Data Center Networks. In 2024 International Conference on Software, Telecommunications and Computer Networks (SoftCOM), 2024.
(6) Won, R. Integrating silicon photonics. *Nat. Photonics* **2010**, *4* (8), 498-499.
(7) Shekhar, S.; Bogaerts, W.; Chrostowski, L.; Bowers, J. E.; Hochberg, M.; Soref, R.; Shastri, B. J. Roadmapping the next generation of silicon photonics. *Nat. Commun.* **2024**, *15* (1), 751.
(8) Rickman, A. The commercialization of silicon photonics. *Nat. Photonics* **2014**, *8* (8), 579-582.
(9) Bogaerts, W.; Perez, D.; Capmany, J.; Miller, D. A. B.; Poon, J.; Englund, D.; Morichetti, F.; Melloni, A. Programmable photonic circuits. *Nature* **2020**, *586* (7828), 207-216.
(10) Shastri, B. J.; Tait, A. N.; Ferreira de Lima, T.; Pernice, W. H. P.; Bhaskaran, H.; Wright, C. D.; Prucnal, P. R. Photonics for artificial intelligence and neuromorphic computing. *Nat. Photonics* **2021**, *15* (2), 102-114.
(11) Mourgias-Alexandris, G.; Moralis-Pegios, M.; Tsakyridis, A.; Simos, S.; Dabos, G.; Totovic, A.; Passalis, N.; Kirtas, M.; Rutirawut, T.; Gardes, F. Y.; et al. Noise-resilient and high-speed deep learning with coherent silicon photonics. *Nat. Commun.* **2022**, *13* (1), 5572.
(12) Shu, H.; Chang, L.; Tao, Y.; Shen, B.; Xie, W.; Jin, M.; Netherton, A.; Tao, Z.; Zhang, X.; Chen, R.; et al. Microcomb-driven silicon photonic systems. *Nature* **2022**, *605* (7910), 457-463.
(13) Chen, G.; Yu, Y.; Shi, Y.; Li, N.; Luo, W.; Cao, L.; Danner, A. J.; Liu, A. Q.; Zhang, X. High‐Speed Photodetectors on Silicon Photonics Platform for Optical Interconnect. *Laser Photonics Rev.* **2022**, *16* (12).
(14) Yang, K. Y.; Shirpurkar, C.; White, A. D.; Zang, J.; Chang, L.; Ashtiani, F.; Guidry, M. A.; Lukin, D. M.; Pericherla, S. V.; Yang, J.; et al. Multi-dimensional data transmission using inverse-designed silicon photonics and microcombs. *Nat. Commun.* **2022**, *13* (1), 7862.
(15) Shi, Y.; Zhang, Y.; Wan, Y.; Yu, Y.; Zhang, Y.; Hu, X.; Xiao, X.; Xu, H.; Zhang, L.; Pan, B. Silicon photonics for high-capacity data communications. *Photonics Res.* **2022**, *10* (9).
(16) Rizzo, A.; Novick, A.; Gopal, V.; Kim, B. Y.; Ji, X.; Daudlin, S.; Okawachi, Y.; Cheng, Q.; Lipson, M.; Gaeta, A. L.; et al. Massively scalable Kerr comb-driven silicon photonic link. *Nat. Photonics* **2023**, *17* (9), 781-790.
(17) Qiao, L.; Tang, W.; Chu, T. 32 x 32 silicon electro-optic switch with built-in monitors and balanced-status units. *Sci. Rep.* **2017**, *7*, 42306.
(18) Cheng, Q.; Dai, L. Y.; Abrams, N. C.; Hung, Y.-H.; Morrissey, P. E.; Glick, M.; O'Brien, P.; Bergman, K. Ultralow-crosstalk, strictly non-blocking microring-based optical switch. *Photonics Res.* **2019**, *7* (2).
(19) Suzuki, K.; Konoike, R.; Matsuura, H.; Matsumoto, R.; Inoue, T.; Namiki, S.; Kawashima, H.; Ikeda, K. Recent Advances in Large-scale Optical Switches Based on Silicon Photonics. In Optical Fiber Communication Conference (OFC) 2022, 2022.
(20) Gao, W.; Li, X.; Lu, L.; Liu, C.; Chen, J.; Zhou, L. Broadband 32 × 32 Strictly‐Nonblocking Optical Switch on a Multi‐Layer Si3N4-on-SOI Platform. *Laser Photonics Rev.* **2023**, *17* (11).
(21) Chen, X.; Lin, J.; Wang, K. A Review of Silicon-Based Integrated Optical Switches. *Laser Photonics Rev.* **2023**, *17* (4).
(22) Li, X.; Gao, W.; Lu, L.; Chen, J.; Zhou, L. Ultra-low-loss multi-layer 8 × 8 microring optical switch. *Photonics Res.* **2023**, *11* (5).
(23) Sato, K.-i. Optical switching will innovate intra data center networks [Invited Tutorial]. *J. Opt. Commun. Networking* **2023**, *16* (1).
(24) Namiki, S. Optical Switching Challenges for the Post-Moore's Law Era. In 50th European Conference on Optical Communications (ECOC 2024), 2024.
(25) Dupuis, N.; Rylyakov, A.; Schow, C.; Kuchta, D.; Baks, C.; Orcutt, J.; Gill, D.; Green, W.; Lee, B. Nanosecond-scale Mach-Zehnder-based CMOS Photonic Switch Fabrics. *J. Lightwave Technol.* **2016**, 1-1.
(26) Chu, T.; Chen, N.; Tang, W.; Wu, Y. Large-Scale High-Speed Photonic Switches Fabricated on Silicon-Based Photonic Platforms. In Optical Fiber Communication Conference (OFC) 2023, 2023.
(27) Soref, R.; Bennett, B. Electrooptical effects in silicon. *IEEE J. Quantum Electron.* **1987**, *23* (1), 123-129.
(28) Dupuis, N.; Lee, B. G.; Rylyakov, A. V.; Kuchta, D. M.; Baks, C. W.; Orcutt, J. S.; Gill, D. M.; Green, W. M. J.; Schow, C. L. Design and Fabrication of Low-Insertion-Loss and Low-Crosstalk Broadband 2 × 2 Mach–Zehnder Silicon Photonic Switches. *J. Lightwave Technol.* **2015**, *33* (17), 3597-3606.
(29) Dupuis, N.; Rylyakov, A. V.; Schow, C. L.; Kuchta, D. M.; Baks, C. W.; Orcutt, J. S.; Gill, D. M.; Green, W. M.; Lee, B. G. Ultralow crosstalk nanosecond-scale nested 2 x 2 Mach-Zehnder silicon photonic switch. *Opt. Lett.* **2016**, *41* (13), 3002-3005.
(30) Lu, Z.; Celo, D.; Mehrvar, H.; Bernier, E.; Chrostowski, L. High-performance silicon photonic tri-state switch based on balanced nested Mach-Zehnder interferometer. *Sci. Rep.* **2017**, *7* (1), 12244.
(31) Bao, P.; Cheng, Q.; Wei, J.; Talli, G.; Kuschnerov, M.; Penty, R. V. Harnessing self-heating effect for ultralow-crosstalk electro-optic Mach–Zehnder switches. *Photonics Res.* **2023**, *11* (10).
(32) Bao, P.; Yao, C.; Tan, C.; Yuan, A. Y.; Chen, M.; Savory, S. J.; Penty, R.; Cheng, Q. Ultra-low-crosstalk Silicon Switches Driven Thermally and Electrically. *ArXiv:2410.00592* **2024**.
(33) Xing, J.; Li, Z.; Yu, Y.; Yu, J. Low cross-talk 2 x 2 silicon electro-optic switch matrix with a double-gate configuration. *Opt. Lett.* **2013**, *38* (22), 4774-4776.
(34) Goh, T.; Himeno, A.; Okuno, M.; Takahashi, H.; Hattori, K. High-extinction ratio and low-loss silica-based 8×8 strictly nonblocking thermooptic matrix switch. *J. Lightwave Technol.* **1999**, *17* (7), 1192-1199.
(35) Soldano, L. B.; Pennings, E. C. M. Optical multi-mode interference devices based on self-imaging: principles and applications. *J. Lightwave Technol.* **1995**, *13* (4), 615-627.
(36) Guo, D.; Chu, T. Compact broadband silicon 3 dB coupler based on shortcuts to adiabaticity. *Opt. Lett.* **2018**, *43* (19), 4795-4798.
(37) Iino, K.; Kita, T. Ultrafast operation of Si thermo-optic switch using differential control method. *Jpn. J. Appl. Phys.* **2022**, *62* (1).
(38) Xu, Q.; Manipatruni, S.; Schmidt, B.; Shakya, J.; Lipson, M. 12.5 Gbit/s carrier-injection-based silicon micro-ring silicon modulators. *Opt. Express* **2007**, *15* (2), 430-436.
(39) Tran, M. A.; Komljenovic, T.; Hulme, J. C.; Davenport, M. L.; Bowers, J. E. A Robust Method for Characterization of Optical Waveguides and Couplers. *IEEE Photonics Technol. Lett.* **2016**, *28* (14), 1517-1520.